\documentstyle[prb,preprint,aps]{revtex}
\begin{document}
\draft
\title{Low-energy charge dynamics in La$_{\text{0.7}}$Ca$_{\text{0.3}}$MnO$_{\text3}$:\\THz time-domain spectroscopic studies}
\author{N. Kida}
\address{Research Center for Superconductor Photonics, Osaka University\\and Core Research for Evolutional Science \& Technology (CREST), Japan Science \& Technology Corporation (JST),\\2-1 Yamadaoka, Suita, Osaka 565-0871, Japan}
\author{M. Hangyo}
\address{Research Center for Superconductor Photonics, Osaka University,\\2-1 Yamadaoka, Suita, Osaka 565-0871, Japan}
\author{M. Tonouchi}
\address{Research Center for Superconductor Photonics, Osaka University\\and Core Research for Evolutional Science \& Technology (CREST), Japan Science \& Technology Corporation (JST),\\2-1 Yamadaoka, Suita, Osaka 565-0871, Japan}
\date{\today}
\maketitle
\begin{abstract}
Direct experimental estimations of the low-energy ($1.5\sim10$ meV) complex dielectric constants spectrum and its temperature variation have been investigated for La$_{0.7}$Ca$_{0.3}$MnO$_3$ thin films using terahertz time-domain spectroscopy. At low temperatures, a clear Drude-term emerges. With increasing temperature, the scattering rate increases, while the plasma frequency decreases, derived both from a simple Drude model. Finally, a Drude-term submerges well below the insulator-metal transition temperature. On the basis of the present results, low-energy charge dynamics are discussed.
\end{abstract}


Hole-doped manganites, general formula $A$$_{1-x}${\it B}$_{x}$MnO$_3$ (where $A$ is a trivalent rare-earth element and $B$ a divalent element such as Sr or Ca), exhibit a rich variety of electronic and magnetic properties and versatile intrigued phenomena such as colossal magnetoresistance (CMR). \cite{YTokura1,YTokura2} One of the most important goal in this field is the achievement of understandings of the anomalous state near the insulator-metal (IM) phase boundary, because the conventional CMR effect appears around the IM transition temperature $(T_{\text{IM}})$. A number of research efforts have been focused on the paramagnetic insulating phase of La$_{0.7}$Ca$_{0.3}$MnO$_3$ and revealed the essential ingredient for the occurrence of the CMR effect; Jahn-Teller small polaron process, \cite{MFHundley,PDai} in addition to the traditionally well known, double-exchange process. On the other hand, the ferromagnetic metallic phase is not well clarified.

In this Communication, low-energy ($1.5\sim10$ meV) charge dynamics of La$_{0.7}$Ca$_{0.3}$MnO$_3$ thin films in the ferromagnetic metallic phase have been investigated by means of terahertz (THz) time-domain spectroscopy (TDS) with transmission configuration.

Optical spectroscopy has given useful information of the electronic structure in the strong correlated electron system. Experimental studies based on temperature-dependent optical conductivity spectra have shown anomalous features of manganites as firstly demonstrated by Okimoto {\it et al}.; the 100\% spin polarized half-metallic nature in the ferromagnetic metallic phase, \cite{YOkimoto1} the large spectral weight change up to several eV with temperature and magnetic field, \cite{YOkimoto1,SGKaplan,YOkimoto2,YMoritomo1,KHKim2,AMachida,MQuijada,AVBoris,ESaitoh,KTakenaka1,KTakenaka2} and a small Drude weight with the incoherent part background. \cite{YOkimoto1,KHKim1,YOkimoto2,ESaitoh,JRSimpson}

The first point is well established directly by spin-polarized photoemission spectroscopy. \cite{JHPark} The second one is understood in terms of the large spectral weight transfer from the interband transition to the exchange-split {\it e}$_{\text g}$-bands, as spin being perfectly polarized. However, last one, low-energy charge dynamics are still far from being understood and controversial under several groups. For example, a sharp Drude-term below the optical phonon energy with nearly $\omega$-independent flat incoherent part have been observed in La$_{1-x}$Sr$_x$MnO$_3$ \cite{YOkimoto1,YOkimoto2} and La$_{0.7}$Ca$_{0.3}$MnO$_3$, \cite{KHKim1,KHKim2} although the estimated effective mass is large compare to one from the specific heat measurement. \cite{TOkuda1,TOkuda2} On the contrary, Takenaka {\it et al}. have claimed that a sharp Drude-term falls on the incoherent part in La$_{1-x}$Sr$_x$MnO$_3$, when the cleaved surface is used. \cite{KTakenaka1,KTakenaka2}

Most experimental demonstrations have been done using Kramers-Kronig (KK) transformation to estimate complex optical spectra. The method of this type suffers from the fact that the low-energy spectrum is deduced from extrapolation using the DC resistivity or Hagen-Rubens relation in order to execute KK transformation. On the contrary, using TDS technique, direct measurements of time profile of the transmission amplitude yields both real and imaginary parts of complex optical spectra without KK transformation. In addition to the above advantage, the THz beam is suitable for the observation of a Drude-term in the metallic phase. They overcome the difficulty of detailed discussions in the low-energy region, which are limited to far-infrared spectroscopy as mentioned above. We carefully analyzed the data of dielectric constant spectra of La$_{0.7}$Ca$_{0.3}$MnO$_3$ thin films by a simple Drude model including the plasma frequency and the scattering rate as parameters. We focus on their temperature dependence in the ferromagnetic metallic phase and discuss the low-energy charge dynamics with a comparison of the various spectroscopic studies.

The sample investigated here is La$_{0.7}$Ca$_{0.3}$MnO$_3$ thin film, deposited on MgO(100) substrate by pulsed laser deposition technique. The obtained film is $a$-axis oriented and has a lattice constant of 3.86 \AA, evaluated from room temperature X-ray diffraction profile. $T_{\text{IM}}$ is about 230 K according to the temperature dependence of the resistivity measurement [see, the inset of Fig. \ref{fig3}(a)]. The surface of the nondoped InAs wafer is excited by the femtosecond laser pulses from a mode-locked Ti:Sapphire laser emitting a 50 fs pulses at 800 nm to produce the THz beam. Transmitted THz radiation through the sample is detected by the bow-tie-type low temperature grown GaAs photoconductive switch. The waveform of the radiation is obtained by scanning the time delay. Namely, we get information of the transmission amplitude as well as phase shift without KK transformation. In order to minimize experimental errors, we repeated the same procedure $3\sim6$ times at respective temperatures.

The TDS geometry of thin film deposited on MgO substrate is shown in Fig. \ref{fig1}. We ignore the multiple reflectance inside MgO. The complex transmission coefficient of MgO can be written as 
\begin{equation}
t^\prime=t_3t_3^\prime\exp[-i(k_2-k_0)d^\prime],
\label{eqn.1}
\end{equation}
where $d^\prime$ is the thickness ($\sim$0.5 mm) of MgO, $t_3=2N_2/(1+N_2)$ is the transmission coefficient from MgO to air, $t_3^\prime=2/(2N_2+1)$ is the transmission coefficient from air to MgO, $k_2$ the wavenumber of MgO, $k_0$ the vacuum wavenumber, and $N_2$ the complex refractive indices of MgO. The complex transmission coefficient of film/MgO can be written as
\begin{equation}
t=t_1t_2t_3\frac{\exp\{-i[k_1d+k_2d^\prime-k_0(d+d^\prime)]\}}{1+r_1r_2\exp(-i2k_1d)},
\label{eqn.2}
\end{equation}
where {\it d} is the thickness ($\sim$45 nm) of film, $t_1=2/(N_1+1)$ is the transmission coefficient from air to film, $t_2=2N_1/(N_1+N_2)$ is the transmission coefficient from film to MgO, $r_1=(1-N_1)/(1+N_1)$ is the reflective coefficient from the interface between air and film, $r_2=(N_1-N_2)/(N_1+N_2)$ is the reflective coefficient from the interface between film and MgO, $k_1$ the wavenumber of film, and $N_1$ the complex refractive indices of film. From eqs. (\ref{eqn.1}) and (\ref{eqn.2}) can be obtained as
\begin{equation}
\frac{t_1t_2}{t_3^\prime}\frac{\exp[-i(k_1-k_0)d]}{1+r_1r_2\exp(-i2k_1d)}=\frac{A}{A^\prime}\exp[i(\theta-\theta^\prime)],
\label{eqn.3}
\end{equation}
where $A$ and $\theta$ are the amplitude and the phase shift of the transmission of film/MgO, respectively. $A^\prime$ and $\theta^\prime$ are those of the transmission of MgO, respectively. $N_1$ can be numerically calculated in eq. (\ref{eqn.3}) using experimentally determined values; $A$, $A^\prime$, $\theta$, and $\theta^\prime$.

Figure \ref{fig2} shows the imaginary part of dielectric constant spectra $[\epsilon_2(\omega)]$ below 10 meV at various temperatures in the ferromagnetic metallic phase. Open circles are experimental data. The power spectrum of the THz source used in this experiment is shown in Fig. \ref{fig2} as the dashed line. Reflecting the metallic character, $\epsilon_2$ shows a steep increase with decreasing photon energy and a Drude peak centered at $\hbar\omega\sim0$ is clearly seen at 16 K. With increasing temperature, the intensity of a Drude-term decreases and is not seen above 170 K ($\sim$$0.7T_{\text{IM}}$). It is noticed that this temperature is well below $T_{\text{IM}}$ measured by the conventional four-probe method as shown in the inset of Fig. \ref{fig3}(a). For quantitative discussions, we applied a simple Drude model. A simple Drude model including two parameters, the plasma frequency $(\omega_{\text p})$ and the scattering rate $(\Gamma)$,
\begin{equation}
\epsilon_2(\omega)=\frac{\Gamma\omega_{\text p}^2}{\omega^2+\Gamma^2}\frac{1}{\omega}
\label{eqn.4}
\end{equation}
has been previously given the fit of the optical conductivity spectrum. \cite{YOkimoto1,SGKaplan,YOkimoto2,KHKim2,AVBoris,ESaitoh,KTakenaka1,KTakenaka2} The solid curves are least-squares fit to experimental data using eq. (\ref{eqn.4}). We found that the dielectric response between 16 and 144 K is successfully explained by a simple Drude model. We obtain $\omega_{\text p}\sim$ 1.6 eV and $\Gamma\sim$ 100 meV at 16 K. The value of $\omega_{\text p}$ is consistent with the previous far-infrared spectroscopic result of La$_{0.7}$Ca$_{0.3}$MnO$_3$ thin film deposited on LaAlO$_3$ substrate by Simpson {\it et al}. \cite{JRSimpson} However, the value of $\Gamma$ is found to be 5-times larger than that of Simpson {\it et al}. ($\Gamma\sim$ 20 meV). At low temperatures, the impurity scattering is the major factor of great influence in $\Gamma$. The value of the residual resistivity is a good measure of the strength of the impurity scattering; our sample has the residual resistivity about 300 $\mu\Omega$cm, which is 3-times larger than the value reported by Simpson {\it et al}. Therefore, this discrepancy mainly originates from the crystal quality between two samples.

To quantify the spectral change with temperature, we plotted in Fig. \ref{fig3}, the temperature dependence of (a) the resistivity, (b) $\Gamma$, and (c) $\omega_{\text p}$ below 200 K. As clearly seen, a Drude-term submerges above 170 K. With increasing temperature, $\Gamma$ increases in proportion to $T^2$ below 160 K, as indicated by the solid line in Fig. \ref{fig3}(b). But the clear deviation of $\Gamma$ from the $T^2$-term can be seen above 160 K. Accordingly, the temperature dependence of the resistivity has the $T^2$-term as described later. This rise of $\Gamma$ has been reported by Simpson {\it et al}. \cite{JRSimpson} However, the $T^2$-coefficient is more pronounced when compared to Simpson {\it et al}. In contrast to the intuitive view of conventional metals, $\omega_{\text p}$ decreases with increasing temperature as shown in Fig. \ref{fig3}(c). It suggests that $m^\ast$ and/or the carrier density have the temperature dependence in the ferromagnetic metallic phase.

It is well known that all the resistance in the ferromagnetic metal is described by the Matthiessen's rule expressed as
\begin{equation}
\rho(T)=\rho_0+AT^2+BT^{9/2},
\label{eqn.5}
\end{equation}
where $\rho_0$ is the residual resistivity, the $T^2$-term the electron-electron scattering process, and the $T^{9/2}$-term the electron-magnon scattering process as introduced by the double-exchange model. \cite{KKubo} The temperature dependence of the resistivity $[1/\sigma_1(0)]$ from the following relationship is represented by closed circles as shown in Fig. \ref{fig3}(a).
\begin{equation}
\omega_{\text p}=\sqrt{\frac{\sigma_1(0)\Gamma}{\epsilon_0}},
\label{eqn.6}
\end{equation}
where $\epsilon_0$ is the permittivity of vacuum. The dashed line is a least-squares fit using eq. (\ref{eqn.5}) to the following data; $\rho_0=3.2\times10^{-4}$ $\Omega\text{cm}$, $A=1.5\times10^{-8}$ $\Omega\text{cm/K}^2$, and $B=1.9\times10^{-13}$ $\Omega\text{cm/K}^{9/2}$. As previously reported, the metallic phase is well described by the first- and second-terms in eq. (\ref{eqn.5}). \cite{TOkuda1,TOkuda2,AUrushibara} We also performed the fit using only $T^2$-term as shown by the solid line. Both lines hold well with the experimental data below 100 K. These result indicates that the scattering process is dominated by only the electron-electron scattering process, and the electron-magnon and/or the lattice contributions are less important in the ferromagnetic metallic phase below 100 K. \cite{Ref1} It is mentioned that the submergence of the Drude-term, characterized by $1/\sigma_1(0)$ and $\Gamma$, coincides with clear deviations above 160 K from fitting curves as shown in Figs. \ref{fig3}(a) and (b).

As a convenience, we classified the electronic state into three categories below 300 K as shown in the inset of Fig. \ref{fig3}(a). Below $\sim$$0.7T_{\text{IM}}$, a clear Drude-term emerges in the ferromagnetic metallic phase (phase-I). We call phase-II between $\sim$$0.7T_{\text{IM}}$ and $T_{\text{IM}}$, in which a Drude-term is not seen, while $\rho$ shows the metallic conduction $(d\rho/dT>0)$. The polaron hopping regime in the paramagnetic insulating phase (phase-III) is widely reported. \cite{MFHundley,PDai,SHChun1} This classification is recently proposed to clear the electronic state of La$_{0.825}$Sr$_{0.175}$MnO$_3$ by Takenaka {\it et al}. \cite{KTakenaka2,KTakenaka3}

To get insights into features of charge carriers in phase-I, it is important to estimate the effective mass $(m^\ast)$. We assume the carrier density $(n)$ of 1.6 holes per Mn-site according to the Hall effect measurement by Chun {\it et al}. \cite{SHChun1} At 16 K, $m^\ast$ is estimated to be $\sim$15 in units of the bare electron mass $(m_0)$, which can be given by the relationship $m^\ast/m_0=e^2n/(\epsilon_0\omega_{\text p}^2m_0)$. This value is few-times larger than one obtained from the specific heat measurement, \cite{TOkuda2} which is simply ascribed to that the motion of charge carriers has the incoherent nature even in the ferromagnetic metallic ground-state. 

It was recently shown that the electronic specific heat constant of La$_{1-x}$Ca$_x$MnO$_3$ linearly depending on $m^\ast$, shows the less mass renormalization near the insulator-metal critical point $(x_{\text c}\sim0.22)$, \cite{TOkuda2} in contrast to other hole-doped transition metal oxides showing the large mass renormalization effect. We also measured the complex optical spectrum near $x_{\text c}$ using THz TDS technique at 20 K. \cite{NKida} Despite the fact that the magnitude of $m^\ast$ derived from THz TDS technique is considerably different from one in specific heat measurements as described above, no significant enhancement of $m^\ast$ is also observed in our spectroscopic studies; we derived $\omega_{\text p}\sim1.3$ eV and obtained $m^\ast\sim$ 23$m_0$ near $x_{\text c}$ in the ferromagnetic metallic ground-state.

These incoherent characteristics of the charge carriers showing a small Drude weight in manganites are firstly reported by Okimoto {\it et al}.; \cite{YOkimoto1,YOkimoto2} to the best of our knowledge, all values of $m^\ast$ in La$_{1-x}$Sr$_x$MnO$_3$ and La$_{0.7}$Ca$_{0.3}$MnO$_3$ reported previously are inconsistent with that estimated from specific heat measurements \cite{TOkuda1,TOkuda2} except for the results of Takenaka {\it et al}. \cite{KTakenaka1,KTakenaka2} They have claimed that a Drude weight is not a small when the cleaved surface is used. On the contrary, recent careful photoemission spectroscopic studies using the cleaved surface revealed that the spectral weight is reduced at Fermi-level due to the formation of the pseudo-gap. \cite{TSaitoh} These controversial results are not clear at present.

Other important physical quantities at low temperatures are the mean free-path $(l)$ and the Fermi wavelength $(\lambda_{\text F})$. We obtain $l\sim4.7$ \AA, via the relation $\l=\hbar/(\Gamma{m^\ast})(6\pi^2n/p)^{1/3}$, where $p(=2)$ is the degeneracy. Assuming the spherical Fermi surface, $\lambda_{\text F}[=(8\pi/3n)^{1/3}]$ is estimated to be $\sim$6.7 ${\text \AA}$. Therefore, the condition $\lambda_{\text F}\sim l>a$ (lattice constant) is derived even in the ferromagnetic metallic ground-state. This is in contrast to the conventional picture of ordinary metals, in which the condition $\lambda_{\text F}<l$ is satisfied. The observed magnitude of $l$ is comparable with that of $\lambda_{\text F}$, which is simply attributed to the small $\omega_{\text p}$. In addition, the value of $l$ is larger than that of $a$. Therefore, the charge carrier easily hops to another sites, but its motion is mainly restricted by the electron-electron scattering process as evidenced by the fact that $\rho(T)$ is well described in eq. (\ref{eqn.5}).

With increasing temperature towards phase-II, $\Gamma$ increases, while $\omega_{\text p}$ decreases as can be seen in Figs. \ref{fig3} (b) and (c), respectively. Namely, the hopping amplitude of the charge carrier decreases. In fact, the reverse condition $l<a$ is obtained in the vicinity of phase-II, \cite{Ref2} indicating that mobile electrons interacting with the localized scattering site, cannot hop to neighbor sites. In accordance with above the scenario, the clear deviation of eq. (\ref{eqn.5}) to experimental data can be seen.

Recently, F\"{a}th {\it et al}. have reported the coexistence of the nanometer-scale metallic domain embedded with the insulating matrix in La$_{0.7}$Ca$_{0.3}$MnO$_3$ by means of the scanning tunneling measurement in the vicinity of $T_{\text{IM}}$. \cite{MFath} In this picture, the ferromagnetic metallic phase serves as the metallic domain in neighbor insulating phases. In our measurements, $\epsilon_2$ slightly increases with decreasing photon energy even at around 180 K as can be seen in Fig. \ref{fig2}. Although this weak structure cannot be explained by a simple Drude model, the apparent metallic conduction exists. One of the suitable examination for these phenomena is the existence of two-phase mixtures, small metallic domains in the insulating phase as recently proposed. \cite{AMoreo,MUehara}

In summary, using terahertz time-domain spectroscopy, we have directly measured the imaginary part of dielectric constants spectra [$\epsilon_2(\omega)$] for La$_{0.7}$Ca$_{0.3}$MnO$_3$ thin films as a function of temperature without Kramers-Kronig (KK) transformation as widely used in far-infrared spectroscopy. By applying a simple Drude model in the low-energy region ($1.5\sim10$ meV), the plasma frequency $(\omega_{\text p})$ and the scattering rate $(\Gamma)$ are obtained. At low temperatures, the charge carrier has an incoherent nature arising from perhaps the small $\omega_{\text p}$. Interestingly, with increasing temperature, the submergence of a Drude-term in spite of the existence of the metallic conduction occurs well below the insulator-metal transition temperature $(T_{\text{IM}})$. This phenomenon is the fascinating subject from view points of the electronic phase separation in colossal magnetoresistive manganites.

We thank Dr. T. Nagashima and Dr. O. Morikawa for fruitful comments on performing the calculation and A. Quema for reading the manuscript.


\begin{figure}
\caption{Schematic representation of the film/MgO geometry. Arrows from left (right) to right (left) represent the direction of the THz beam.\label{fig1}}
\end{figure}

\begin{figure}
\caption{Imaginary part of dielectric constant spectra as a function of temperature in the ferromagnetic metallic phase. Solid lines represent the fit using eq. (\ref{eqn.4}) in the text. The dashed line shows the power spectrum of the light source.\label{fig2}}
\end{figure}

\begin{figure}
\caption{Temperature dependence of (a) the resistivity estimated from eq. (\ref{eqn.6}), (b) the scattering rate, and (c) the plasma frequency below 200 K. Solid lines denote fitting results, assuming the $T^2$-dependence. Dashed line represents the fit using eq. (\ref{eqn.5}) in the text. Inset of (a) shows the logarithmic scale of the temperature-dependent resistivity measured by the four-probe method below 300 K. For the classification of vertical lines, see the text.\label{fig3}}
\end{figure}


\begin{references}
\bibitem{Ref1} We also performed the calculation using eq. (\ref{eqn.5}) add to the $T^5$-term the lattice vibration process, because the Debye temperature is approximately 400 K. \cite{TOkuda2} However, the negligible difference can be seen below 100 K. Therefore, we excluded the $T^5$-term in eq. (\ref{eqn.5}).
\bibitem{NKida} Using pulsed laser deposition technique, we prepared the oxygen-deficient $(\delta)$ thin film of La$_{0.7}$Ca$_{0.3}$MnO$_{3-\delta}$ at the reduced oxygen condition. The temperature-dependent resistivity measurement clearly shows that the obtained film has the IM transition around 190 K. According to the phase diagram, \cite{TOkuda2} our sample locates near the IM phase boundary. In order to obtain $m^\ast$ in this material, we assume $n\sim1$ holes/Mn-site.
\end{references}
\end{document}